\documentclass{PoS}

\usepackage[authoryear,square]{natbib}
\bibpunct{(}{)}{;}{a}{}{,}

\usepackage{psfig}
\usepackage{graphicx}

\title{Euclid \& SKA Synergies}

\ShortTitle{Euclid \& SKA Synergies}

\author{\speaker{Thomas D. Kitching}$^{1}$, David Bacon$^2$, Michael L. Brown$^3$, Philip Bull$^4$, 
Jason D. McEwen$^1$, Masamune Oguri$^5$, Roberto Scaramella$^6$, Keitaro Takahashi$^7$, Kinwah Wu$^1$, Daisuke Yamauchi$^8$
\\
$^1$University College London, Department of Space and Climate Physics;
$^2$ICG, University of Portsmouth;
$^3$Department of Physics, University of Manchester;
$^4$University of Oslo, Institute for Theoretical Astrophysics; 
$^5$Graduate School of Science, University of Tokyo, Japan; 
$^6$INAF-Osservatorio di Roma, I-00040 Monteporzio Catone, Italy; 
$^7$Graduate School of Science and Technology, Kumamoto University, Japan; 
$^8$Research Center for the Early Universe, University of Tokyo, Japan;
\\
E-mail: \email{t.kitching@ucl.ac.uk}
}

\abstract{Over the past few years two of the largest and highest fidelity experiments conceived have been approved for 
construction: Euclid is an ESA M-Class mission that will map three-quarters of the extra galactic sky with Hubble Space Telescope resolution 
optical and NIR imaging, and NIR spectroscopy, its scientific aims (amongst others) are to create a map of the dark Universe and to determine the nature 
of dark energy. The Square Kilometre Array (SKA) has similar scientific aims (and others) using radio wavelength observations. 
The two experiments are synergistic in several respects, both through the scientific objectives and through the control of systematic effects. 
SKA Phase-1 and Euclid will be commissioned on similar timescales offering an exciting opportunity to exploit synergies between these facilities.}

\FullConference{
Advancing Astrophysics with the Square Kilometre Array\\
June 8-13, 2014\\
Giardini Naxos, Italy}

\newcommand{\skipthis}[1]{}

\begin{document}

\section{Introduction}
In this Chapter we will describe the ESA Euclid mission, and discuss some of the synergies that Euclid has with the SKA. 
Euclid probes the low redshift Universe, through both weak lensing and galaxy clustering measurements. The SKA has the 
potential to probe a higher redshift regime and a different range in scales of the matter power spectrum, linear scales 
rather than the quasi-non-linear scales that Euclid will be sensitive to, that will make the combination particularly 
sensitive to signatures of modified gravity and neutrino mass. In combination a longer baseline in redshift will improve 
expansion history and growth of structure measurements
leading to improved measurements of the redshift behaviour of the dark energy equation of state. The cross correlation 
between Euclid and SKA weak lensing (shape as well as size and flux magnification), Euclid and SKA galaxy clustering (BAO and redshift space distortions), 
SKA 21cm intensity mapping and Planck data (CMB, SZ and ISW) will provide a multitude of cross-correlation statistics. 
 
As an example of the benefit of combining the experiments for systematic control for the primary science, the shape measurement 
from Euclid and SKA will be affected by systematics in different ways, meaning that a cross-correlation of the weak lensing data 
from both data sets will be less prone to shape measurement biases (see Brown et al., in this volume). Furthermore the intrinsic (un-lensed) ellipticity 
alignments measured by both Euclid and the SKA will have mutual benefit. Specifically, in addition to Euclid, for weak lensing 
SKA contributes i) depth, providing more source counts for weak lensing, ii) a way to cross-calibrate shape systematics, 
iii) a way to cross-calibrate intrinsic alignments, and iv) precise redshifts through 21cm line observations. And possibly 
the use of polarisation and 21cm rotational velocities for intrinsic shapes and lensing tomography of high redshift 21cm fluctuations.

In addition to the primary science objectives of Euclid there is a plethora of additional legacy and cosmology synergies. We can 
expect $>10^5$ strong lens detections from Euclid and SKA, 
many with redshifts for the lenses and sources and high-resolution images. The combination of 
Euclid and SKA will better investigate the obscured cosmic star formation history as a function of redshift and environment 
thanks to their combination of areal coverage and sensitivity. For population III, hypernovae and Type II supernovae NIR emission 
will be detectable by Euclid from supermassive Pop III supernovae in dense $10^7$-$10^8$ $M_{\odot}$ haloes out to 
$z = 10 - 15$,  and radio synchrotron emission from Pop III supernova remnants detectable by SKA out to $z = 20$. 

A further synergistic aspect is that each of these areas will require post-operation infrastructure development in terms of a 
very large number of hydrodynamical and N-body simulations to understand the effects of non-linear clustering and baryonic feedback. 
Finally the data analysis of giga-scale catalogues, peta-bytes of data, and post-operation simulations require astrostatistical 
and astroinformatics synergies for astronomers to access and visualise these data sets. 

\subsection{Euclid Overview}
Euclid\footnote{http://euclid-ec.org} is one of the European Space Agency's (ESA) medium (M) class missions that has been selected as part of ESAs 
``cosmic visions'' programme. Euclid was selected by ESA 
in October 2012, to take the second of the M-class mission places, M2, which means that there is a scheduled launch date 
for 2020. The science objective of Euclid 
is primarily to determine the nature of the the phenomenon that is causing the expansion rate of the Universe to 
accelerate; so-called `dark energy'. However the top-level 
science objectives are in fact four-fold and cover all current major open questions in cosmology: 
\begin{itemize} 
\item 
Dynamical Dark Energy: Is the dark energy simply a cosmological constant, or is it a field that evolves dynamically with the expansion of the Universe?
\item 
Modification of Gravity: Alternatively, is the apparent acceleration instead a manifestation of a breakdown of General Relativity on the largest scales, or a failure of the cosmological assumptions of homogeneity and isotropy?
\item 
Dark Matter: What is dark matter? What is the absolute neutrino mass scale and what is the number of relativistic species in the Universe?
\item 
Initial Conditions: What is the power spectrum of primordial density fluctuations, which seeded large-scale structure, and are they described by a Gaussian probability distribution?
\end{itemize} 
More quantitatively the objective of Euclid is to achieve a dark energy Figure of Merit \citep{detf} 
of $\geq 400$ and to determine the growth index $\gamma$, that 
provides a simple parameterisation for deviations from general relativity, to an accuracy of $0.02$. 
Euclid is designed to achieve these science objectives with support from ground-based observations. 
In combination with the SKA it is expected that the combined strength of these experiments will enable both to far exceed their 
singular goals. 
In order to achieve these science objectives Euclid will use two primary cosmological probes: weak lensing and galaxy clustering. 
These are given equal priority in the mission and indeed it is only through the combination of weak lensing and galaxy clustering that the science objectives can be achieved. 
In addition Euclid will provide the astronomy community with a rich data set that will enable many astrophysical studies; areas in Euclid that are referred to as `Legacy' science. 

The SKA science goals are synergistic with those of Euclid. Whilst Euclid is a focussed experiment to address a particular goal, the SKA design allows for flexibility in 
the type of science questions that can be addressed. In this case proposals for SKA observations can be tailored to be in the best synergy with Euclid. 
The Euclid mission has been designed given particular constraints on the telescope size and overall cost of the mission and instrumentation. An optimisation during the 
Euclid Assesment Phase \citep{yb} that was refined and elaborated during the Definition Phase \citep{rb} of the mission. This resulted in a nominal 
design of Euclid being a $1.2$ meter Korsch, 
$3$ mirror anasigmat telescope. There will be two instruments on board: the VISible focal plane instrument (VIS) \citep{cropper10} and the Near Infrared 
SpectroPhotometric (NISP) instrument. 

VIS will provide high-resolution optical imaging over a field of view of size $0.787\times 0.709$ square degrees, with a resolution of $0.18$ arseconds, over a single 
broad-band wavelength range of $550$-$900$ nm (an optical band equivalent to an RIZ filter, although Euclid in fact does not have an optical filter on board). It will consist of 
$36$ $4$k$\times 4$k CCDs. The primary 
purpose of the VIS instrument is to provide imaging to enable the measurement of galaxy ellipticities to sufficient accuracy and precision 
for use in weak lensing. It will image approximately $1.5$ billion galaxies with a limiting R-band magnitude of $24.5$ ($10\sigma$ extended source). 

NISP has a field of view of a similar size to VIS of $0.763\times 0.722$ square degrees, these are matched to enable 
simultaneous/matched observations of the sky. The focal 
plane will contain $16$ $2$k$\times 2$k HgCdTe (``Mer-Ca-Tel'') detectors. 
There will be $3$ NIR filters, that will enable imaging, and $2$ grism spectroscopic elements, that will 
enable slitless spectroscopy (with an approximate resolution of $R=250$). The imaging is designed to 
enable photometric measurements of the same galaxies observed using the VIS 
instrument for the determination of photometric redshifts for weak lensing. The spectroscopy is designed to enable precision redshifts to 
be determined for galaxy clustering measurements. 

In order to achieve the photometric redshift required for the weak lensing science Euclid will use ground-based optical imaging data from DES and KiDS, and any 
other available surveys; requiring normal broad-band imaging over the VIS wavelength range. Spectroscopic redshifts from ground-based surveys such as BOSS, DESI and MOONS 
will also be ingested and used for photometric redshift calibrations. 

Every aspect of Euclid is designed using a systems engineering approach where the science requirements are translated into progressively more detailed requirements on 
survey, telescope and instrument design, as well as requirements on the algorithms used in the data processing. This flow of requirements is described in a series of 
ESA documents (for example the Euclid SciRD) and related publications (for example \citep{cropper13}). 
With regard to instrument designs, Euclid and the SKA are therefore synergistic in the sense that Euclid will observe in the optical and NIR wavelengths 
and SKA will observe at 
longer radio wavelengths. As we will discuss in this article this enables unique scientific synergies to be exploited when combining the data between the two experiments. 

\subsection{Survey Synergies} 
\begin{figure}
\psfig{file=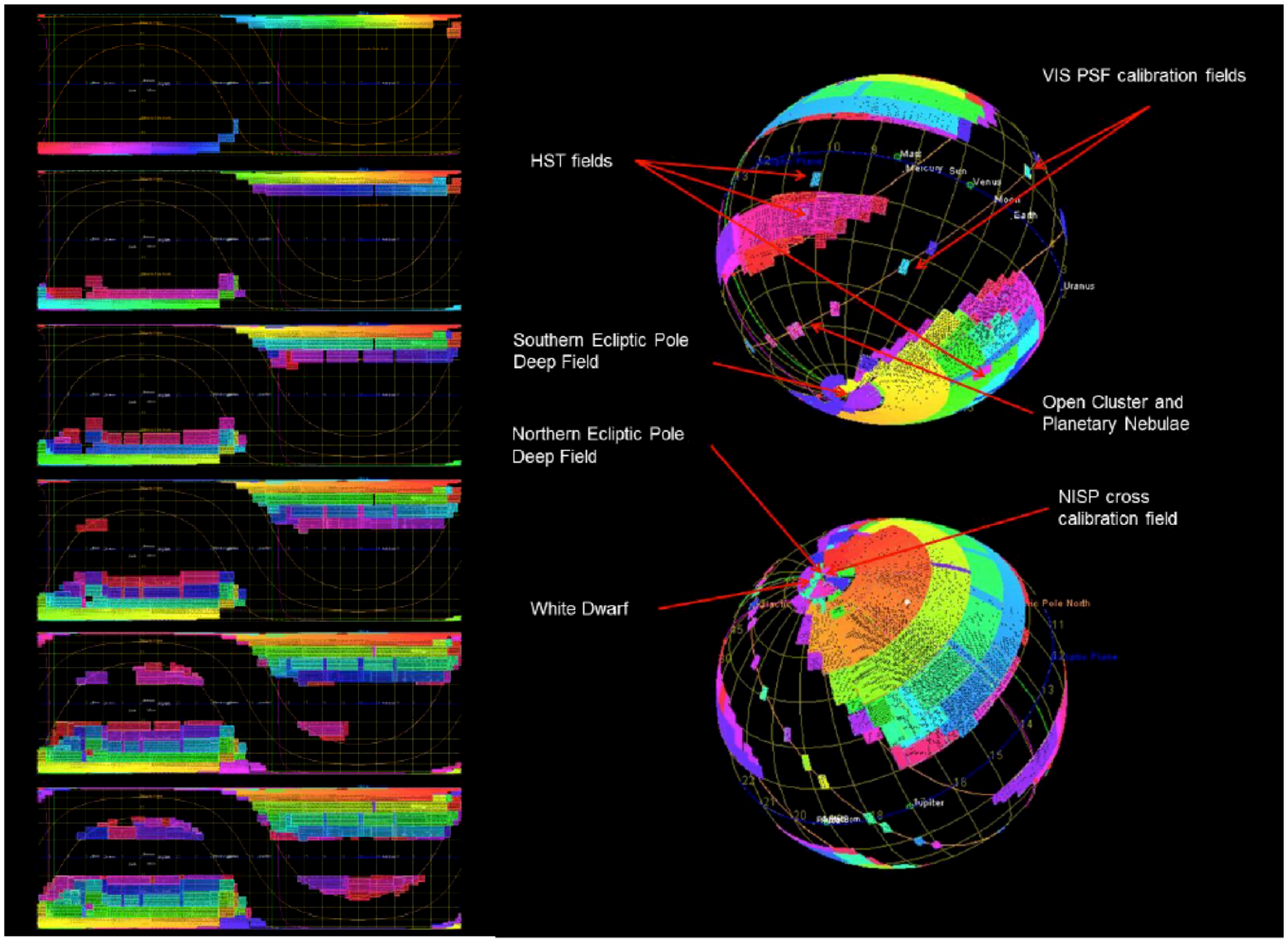,width=\columnwidth,angle=90}
\caption{From \cite{ja}. Euclid Reference Survey construction from year 1 (top) to year 6 (bottom) in cylindrical (left) and orthographic
projection. In the left panel, only the wide survey fields are shown. In the right panel, the calibration fields are also displayed.}
\label{esurvey}
\end{figure}

The Euclid primary probes, weak lensing and galaxy clustering, will be carried out over 
the same area of sky in a wide-field survey of $15$,$000$ square degrees. 
As described in \cite{rb} the area is driven by an optimisation between the galaxy clustering (that prefers a wider area for a fixed observing time) and weak lensing 
(that can prefers a shallower deeper survey for intrinsic alignment mitigation), as well as efficiency in survey design whereby the lowest $30$ degrees in ecliptic 
latitude are not observed to avoid zodiacal light contamination. In Figure \ref{esurvey} 
we show the Euclid reference survey from \cite{ja} and how this builds as a function of time. It is self-evident that the areal coverage of Euclid and
SKA is synergistic with both experiments expected to observe significant fractions of the extragalactic sky. 
The synergy of the Euclid and SKA experiments in time is also self-evident. Euclid is a nominally 6 year mission that is scheduled to 
begin observations in 2020, SKA is scheduled to begin construction in 2018 and begin observations on a similar time scale to Euclid. 

\begin{figure}
\center
\psfig{file=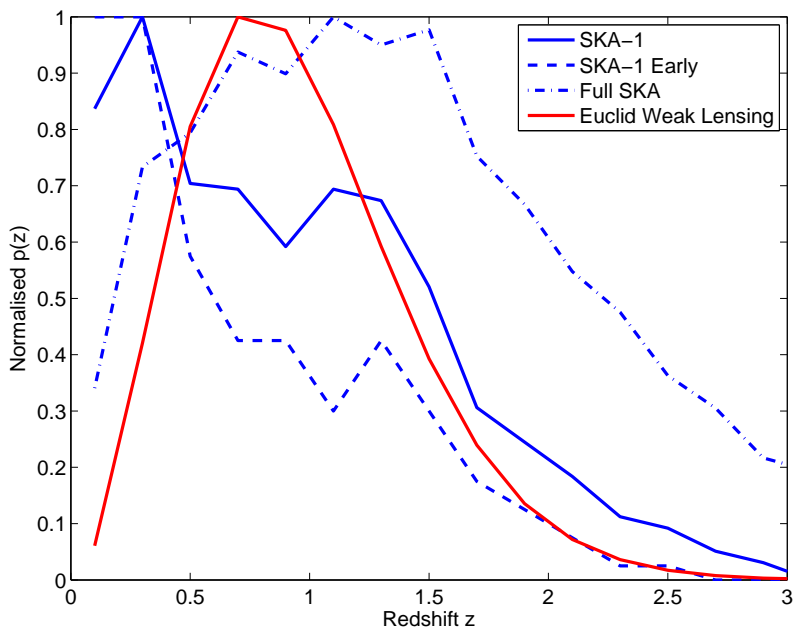,width=0.7\columnwidth}
\caption{The normalised distributions of galaxies usable for weak lensing analyses in Euclid and the SKA, for the different 
phases. The numbers for SKA are derived from specifications quoted in the SKA Imaging Science Performance memo which 
is in turn based on the SKA-1 Baseline Design and from the SKADS simulations of \cite{wilman08}. The instrument of choice is SKA-MID and it has been assumed 
Band 2 (950-1760 MHz) has been used which, under the current version of the SKA-1 baseline design gives 
the best performance in terms of sensitivity at the required angular resolution.}
\label{nz}
\end{figure}

The depth of the Euclid and SKA surveys is also synergistic, Euclid will have a limiting magnitude in optical wavelengths that will mean 
it is sensitive to galaxies with redshifts over the range $0<z<2$, and the NIR spectroscopy will be sensitive to galaxies 
at slightly higher redshifts. In Figure \ref{nz} we show as an example the normalised distribution of weak lensing galaxies that 
the SKA phase 1 and the full SKA, over $3\pi$ steradians, and the Euclid weak lensing survey will cover. 
The Euclid weak lensing survey will have $30$ galaxies per square arcminute with size larger than $1.5$ times the PSF ($R^2$) 
and magnitude $RIZ\leq 24.5$. The SKA survey area and number density has several possibilities, with various area and redshift overlap 
scenarios (see continuum survey overview, also Brown et al. and Jarvis et al. in this volume), 
for example for the full SKA (or SKA phase 2) there could be either
\begin{itemize}
\item 
$3\pi$ steradian survey: RMS noise $= 0.5$ micro-Jy, $10$ gals arcmin$^{-2}$. 
\item 
$5000$ sq. deg survey: RMS noise $= 0.2$ micro-Jy, $23$ galaxies arcmin$^{-2}$. 
\item 
$1000$ sq. deg survey: RMS noise $= 0.1$ micro-Jy, $37$ galaxies arcmin$^{-2}$. 
\end{itemize}
In the remainder of this article we will discuss the scientific synergies that instrument and survey designs of Euclid and the SKA enable.

\section{Cosmological Synergies}
The primary objectives of Euclid are cosmological in nature, in synergy with SKA these objectives can be supplemented and extended. 
Both the SKA and Euclid will be able to measure: 
\begin{itemize}
\item 
Weak lensing shear $g$, 
\item 
Weak lensing magnification; including size $s$ and flux ($n(z)$), 
\item 
Galaxy positions, both photometric $\theta_p$, and spectroscopic $\theta_s$. 
\end{itemize}
This results in $10$ 2-point observables, power spectra, in total that can be combined combinatorially to produce $55$ cross and auto-correlation 
statistics. If, for example, a `tomographic' approach is pursued and the galaxy populations are split into $>10$ redshift bins 
then this results in over $5000$ power spectra that could be computed including all inter and intra-bin correlations. 
Some of these correlations can be used to reduced systematic effects, for example the correlation of galaxy shear with 
galaxy position can be used in infer the intrinsic alignment systematic in cosmic shear studies, see for example \cite{bj10}, 
or the correlation 
between weak lensing shear in the radio and optical data sets can be used to reduce ellipticity measurement systematics. Some 
of these correlations can be used to increase the statistical precision of the combined data, for example the combination 
of weak lensing shear and size \citep{heavens13}. The SKA can also measure unresolved 21cm intensity to create maps that can 
then be included as an additional cosmological probe to be included in the synergistic combination of probes. 

We will highlight just a few of these combinations as examples in this chapter. 

\subsection{21 cm Intensity Mapping}
The principal advantage of 21cm intensity mapping (IM; see Santos et al., in this volume) over traditional galaxy redshift surveys is that extremely large volumes can be 
surveyed in a relatively short time; for example, Phase I of the SKA will be capable of surveying a total area of $\sim$30,000 deg$^2$ 
from $z = 0$ to 3 over the course of 1-2 years. This ability stems from the modest resolution requirements of the IM method; there is 
no need to resolve individual galaxies, and only the integrated HI emission on comparatively large angular scales matters.
We refer to the Chapters on Intensity Mapping and RSD with SKA for more details of this science synergy.

At its most basic, a large IM survey with a Phase I SKA array would complement Euclid simply by increasing the total volume being 
probed. Consider a situation in which Euclid and the SKA targeted independent survey volumes. The sensitivity of SKA-IM to the first 
BAO acoustic peak would be comparable to that of Euclid, which is practically cosmic variance-limited at these scales anyway; doubling 
the survey volume would therefore have a significant effect in beating down cosmic variance and increasing the precision of BAO 
distance indicators and other observables.

Preventing the surveys from overlapping would spoil a number of interesting opportunities, however. Probing the same volume with 
two almost cosmic variance-limited experiments is not redundant; as well as providing useful cross-checks for 
consistency between the two, one can also benefit from the `multi-tracer' effect (see Section \ref{mtm}), whereby one can continue to gain information about 
some observables in spite of cosmic variance as long as different populations of tracers can be distinguished by each survey. This 
can be used to enhance the precision of redshift space distortion measurements, for example, which probe the growth of structure.

\begin{figure}
\center
\psfig{file=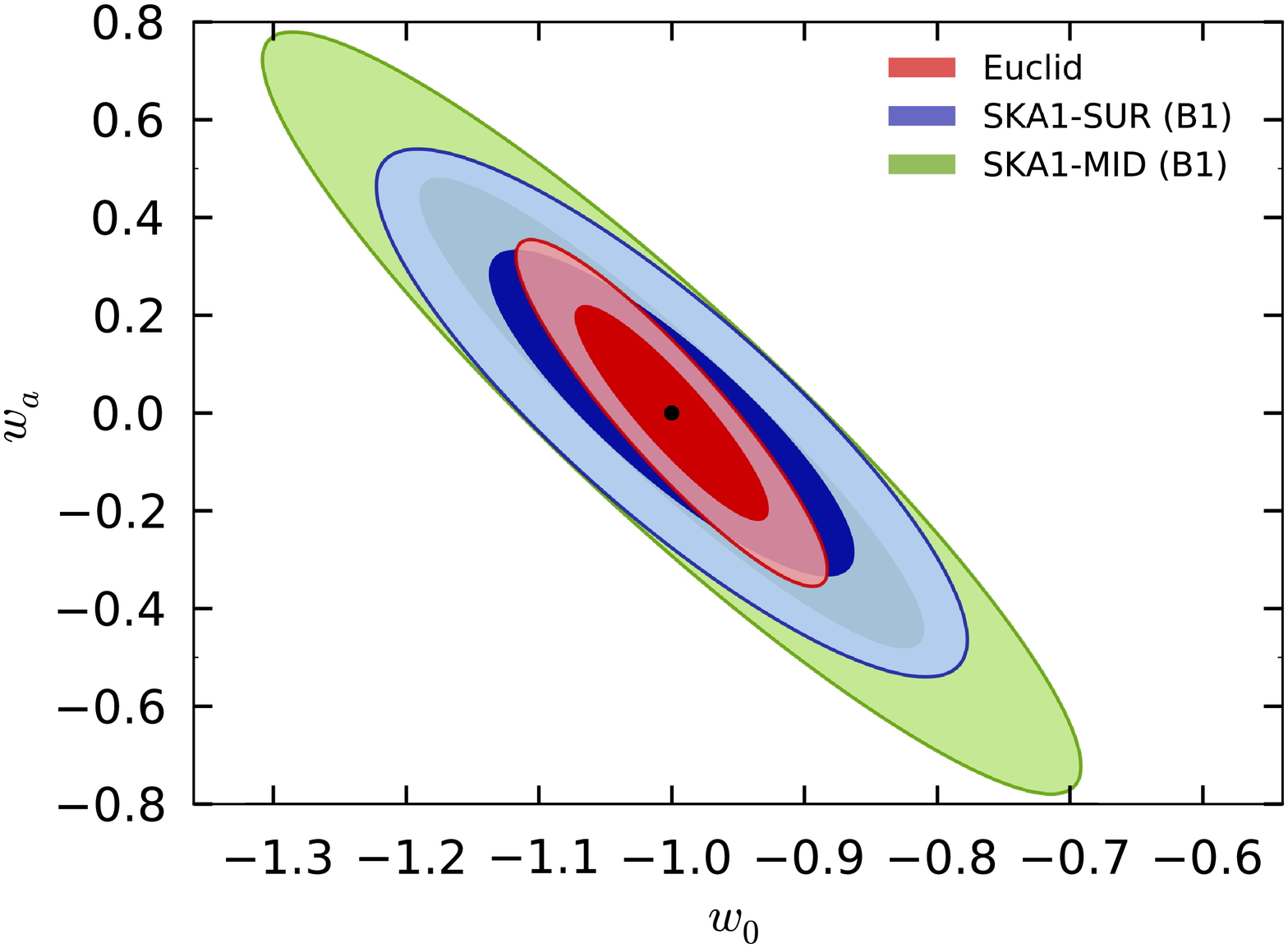,width=0.65\columnwidth,angle=90}
\psfig{file=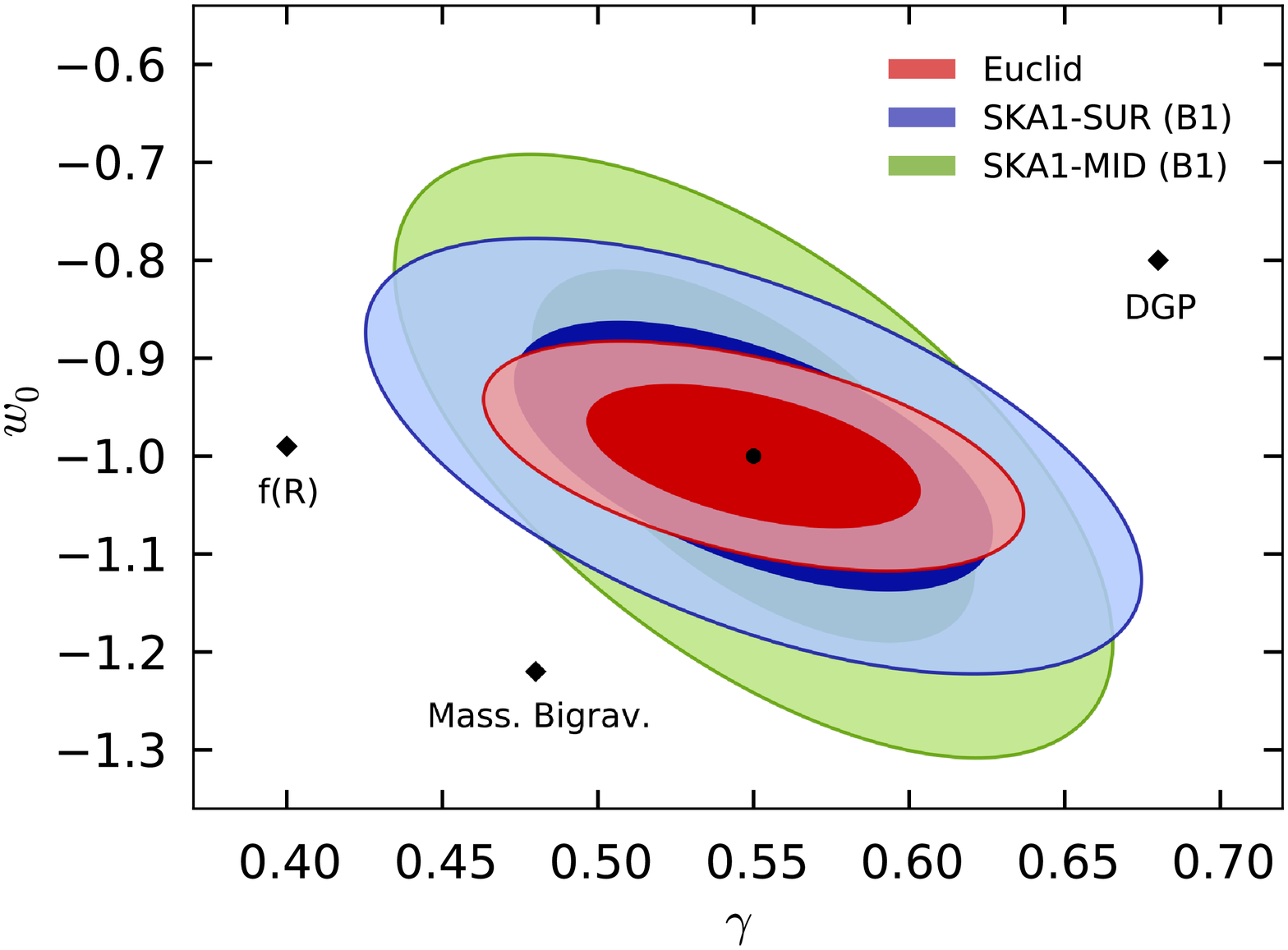,width=0.65\columnwidth,angle=90}
\psfig{file=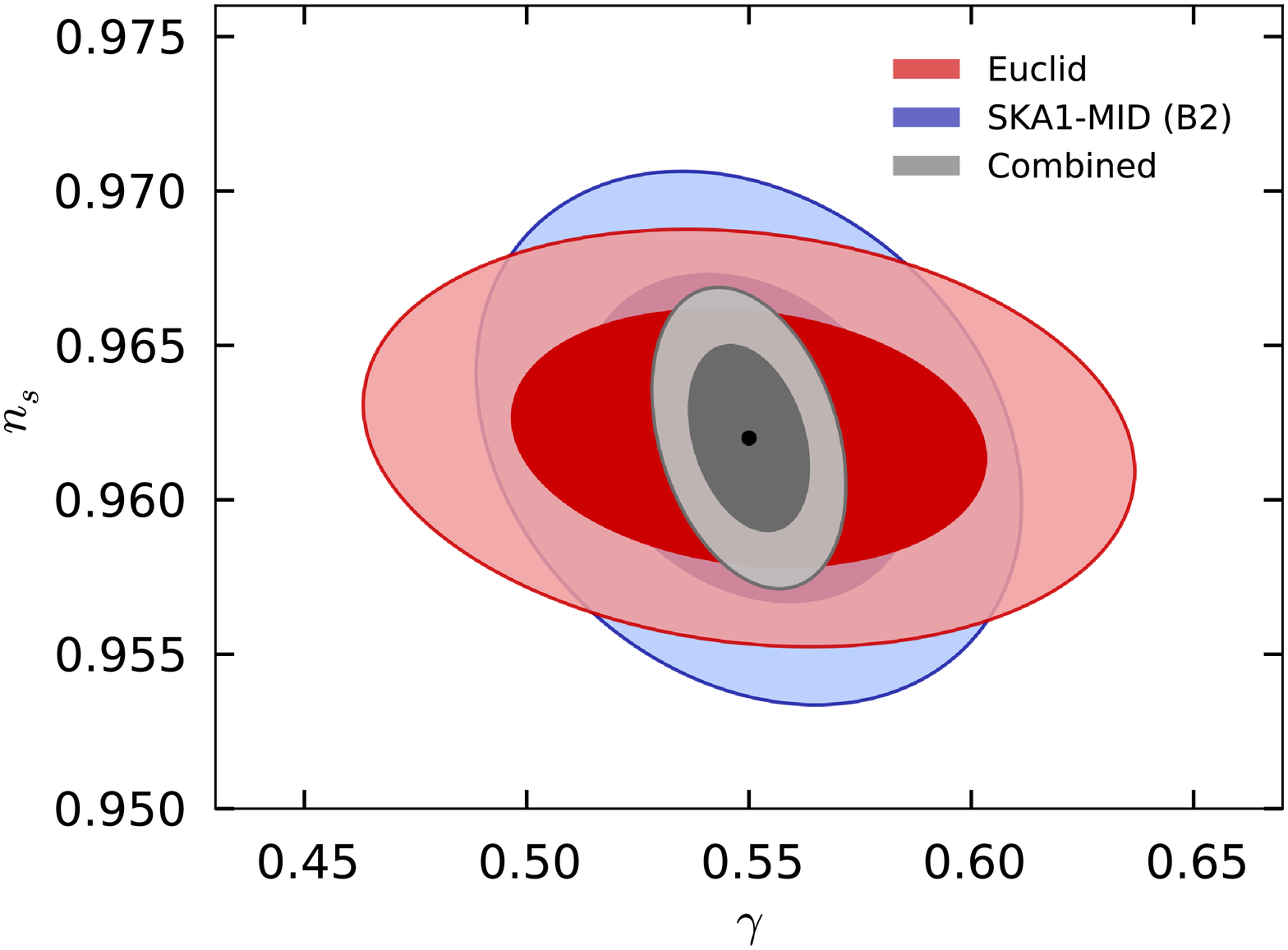,width=0.65\columnwidth,angle=90}
\caption{Predicted $1\sigma$, 2 parameter, error contours for dark energy ($w_0$, $w_a$), modified gravity ($\gamma$), and 
initial condition ($n_s$) parameters for Euclid galaxy clustering and SKA phase 1 design.}
\label{ecosmo}
\end{figure}

SKA IM surveys are capable of significantly extending the redshift range of Euclid, which covers $0<z< 2$. By `filling in' 
redshifts missed by the galaxy survey, one can gain a great deal of leverage on key cosmological functions such as the equation of 
state of dark energy and the linear growth rate; and could also be used for cross-correlations to get better photometric redshifts for Euclid. Figure \ref{ecosmo} 
shows the joint constraints that can be achieved on $w_0$ and the growth index, $\gamma$, 
through the combination of Euclid and an SKA-IM survey with much wider redshift coverage. The additional information at low redshift helps to 
pin down the evolution of dark energy and possible deviations from the standard growth history, while constraints at higher redshift 
act as a useful `anchor', by locating the transition from the matter-dominated era and putting limits on the spatial curvature, $\Omega_K$.

\subsection{Neutrino Physics}
Measuring neutrino properties is potentially a particularly interesting area of Euclid-SKA science. 
Massive neutrino suppress the growth of structure via free-streaming effects 
introducing an effective pressure that acts in regions of higher density. This continues until the 
neutrinos become non-relativistic. This transition imprints a feature at a particular scale in 
the matter power spectrum. As shown in \cite{jim10} an all-sky Euclid-like survey combined with 
a galaxy clustering survey, such as that could be achieved by the final SKA, may even be able to 
determine the neutrino hierarchy.

\subsection{Multi-Tracer Method}
\label{mtm}
The SKA and Euclid will observe a huge number of galaxies, and the errors in
power spectrum of galaxies will be dominated by cosmic variance, rather than
shot noise, at cosmological scales. This is especially serious when we try
to constrain primordial non-Gaussianity whose effect is stronger at larger
scales. Cosmic variance could be avoided with a the `multi-tracer' method 
\cite{sel09} which uses multiple tracers of the dark matter distribution with
different biases to cancel out sample variance. Although power spectra of 
tracers themselves are limited by
cosmic variance, the ratio of the power spectra of two tracers, which
represents the relative bias, can evade cosmic variance and is limited only
by shot noise. Because the mass and redshift dependences of bias are
affected by non-Gaussianity ($f_{\rm NL}$), it can be constrained by the
measurements of relative biases.

\begin{figure}[tp]
\begin{center}
\psfig{file=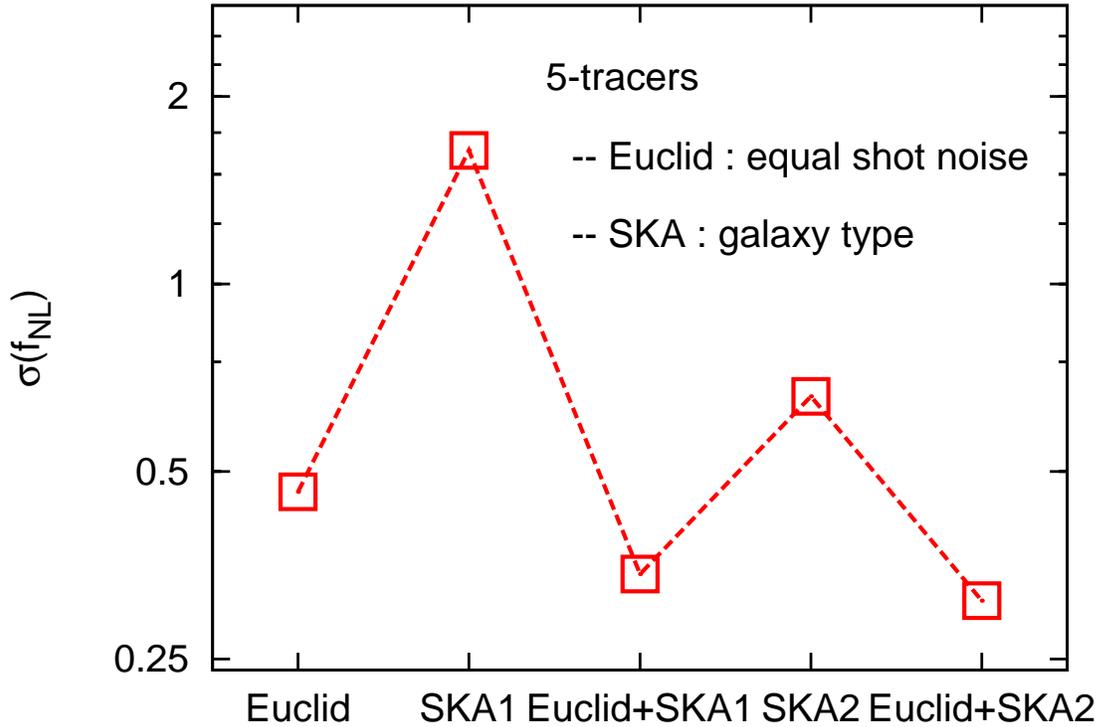,width=\columnwidth,angle=0}
\caption{Expected constraints on $f_{\rm NL}$ from the SKA continuum survey,
Euclid photometric survey and their combinations \cite{yam14}.}
\label{fig:fNL}
\end{center}
\end{figure}

This multi-tracer method is effective when the bias difference, hence mass
difference, is large between tracers and it is critically important to
estimate the mass of the dark matter halo hosting each galaxy. A deep survey is also
important because bias evolves rapidly with redshift. The SKA and Euclid
surveys will have different redshift-distributions of observed galaxies so
that their combination enhances the power of multi-tracer method.
\cite{yam14} studied the potential of combination of the SKA continuum
survey and Euclid photometric survey for the constraint on $f_{\rm NL}$. The
SKA continuum survey reaches much further than the Euclid photometric survey, providing a larger redshift range in combination, 
while the number of galaxies observed by Euclid is larger than that by the
SKA at low redshifts, so they are complementary to probe the evolution of
bias. Figure \ref{fig:fNL} shows expected constraints on $f_{\rm NL}$ from
Euclid, SKA phase 1, the full SKA and their combinations. Here, it is assumed that galaxies
observed by Euclid have photometric redshifts while SKA cannot obtain
redshift information (this is a very conservative assumption, 
but see \citep{sc14} for gains made by using optical and near IR photometric redshifts). 
It is seen that the constraint on $f_{\rm NL}$ can
reach below unity and approach $O(0.3)$.

\subsection{Magnetic Fields} 
The two pressing questions regarding cosmic magnetism are: what is the origin of the 
cosmological magnetic fields and how do these magnetic fields affect (or were affected by) 
the structure formation and evolution of the Universe?
In a more practical perspective, we would like to know how the cosmological magnetic 
fields imprint observational signatures and how we can make use of these signatures to 
better understand the structural formation process in our Universe.

Charged particles in the presence of a magnetic field experience a Lorentz force.
As such the presence of a primordial magnetic field
would produce and amplify density fluctuations in ionised gas (the baryons).
Through gravitational coupling between the baryons and the dark matter, 
the magnetic field would imprint signatures in the dark matter density distributions.
As magnetic fields alter the matter density power spectrum and the halo mass function,
the magnification bias and the redshift distribution of galaxies are distorted 
(see \cite{fm12,sc14}). The capability of Euclid weak lensing studies (and SKA weak lensing studies) 
together with the Faraday rotation measurements using the SKA 
will be able to quantify such a mechanism,
which will not only set constraints on the strength of the magnetic fields in the early Universe 
but also determines the role of magnetic interaction in cosmological structure formation, 
which has been assumed to be solely gravitational.

Gravitational lensing, both weak and strong, are powerful tools in the studies of cosmological large-scale structures.
Lensing calculations rely on ray tracing in a geometrical framework set using the theory of relativity, 
however absorption and scattering of photons emitted from the distant lensed sources along the ray are usually ignored.
This may well be a sensible first approximation, 
but if there are ongoing activities, such as merging of clusters and/or violent large-scale galactic outflows, 
the situation could be more complicated.
Photons on different rays originating from a lensed source could therefore have different frequency dependent attenuation, 
through absorption and/or scattering by the matter along the rays.
Moreover, if there is a spatially inhomogeneous magnetic field permeating the lens and its surrounding, 
polarisation fluctuations could also be induced in the lensed images.
This potential frequency-dependent distortion has not be investigated in detail, however 
it could be inferred through multi-wavelength polarisation observations using the SKA.
With the information of the matter distribution and the magnetic field along the line of slight, 
frequency-dependent attenuation of photons can be modelled using radiative transfer calculations 
and hence prescriptions for correcting the corresponding distortion and biases could be derived.

\section{Systematic Synergies} 
The combination of power spectra discussed in the previous section will enable a mitigation 
systematic to be performed at the statistical level. However because Euclid and the SKA 
will observe many of the same galaxies one can directly compare the measurements to gain an 
ever larger advantage. We refer here to systematic reductions discussed in the Chapter on Weak Lensing and 
Synergies, and highlight here one particular aspect that is enabled through the weak lensing shape 
measurement that is only possible using both Euclid and SKA, due to the PSF stability and size 
available from space or very wide aperture radio telescopes (see \cite{mass12}; such observations 
are not possible using ground-based optical telescopes). 

Optical and radio surveys, such as Euclid and SKA, have a particularly useful synergy in reducing and 
quantifying the impact of systematic effects which may dominate each survey alone on some scales. 
By cross-correlating the shear estimators from one of these surveys with those of the other, several systematic errors are mitigated.
We can see this by writing the contributions to an optical (o) or radio (r) shear estimator:
\begin{eqnarray}
\gamma^{(o)}&=&\gamma_{\rm grav} + \gamma^{(o)}_{\rm int} + \gamma^{(o)}_{\rm sys}\\
\gamma^{(r)}&=&\gamma_{\rm grav} + \gamma^{(r)}_{\rm int} + \gamma^{(r)}_{\rm sys}\\
\end{eqnarray}
where $\gamma_{\rm grav}$ is the gravitational shear we are seeking, $\gamma_{\rm int}$ is the intrinsic ellipticity of the object, and $\gamma_{\rm sys}$ are systematic errors induced by the telescope.
 
If we correlate optical shears with optical shears, or radio shears with radio shears, we obtain terms like
\begin{equation}
\langle \gamma \gamma \rangle = \langle \gamma_{\rm grav} \gamma_{\rm grav} \rangle + \langle \gamma_{\rm grav} \gamma_{\rm int} \rangle + \langle \gamma_{\rm int} \gamma_{\rm int} \rangle + \langle \gamma_{\rm sys} \gamma_{\rm sys} \rangle
\end{equation}
where the first term is the gravitational signal we seek, the second term is the GI intrinsic alignment, the third term is the II intrinsic alignment, and the final term is the contribution from systematics. All of these terms could be similar size on certain scales, which is of damage to cosmological constraints.
On the other hand, if we cross-correlate the optical shears with radio shears, we obtain
\begin{equation}
\langle \gamma^{(o)} \gamma^{(r)} \rangle = \langle \gamma_{\rm grav} \gamma_{\rm grav} \rangle + \langle \gamma_{\rm grav} \gamma^{(o)}_{\rm int} \rangle + \langle \gamma_{\rm grav} \gamma^{(r)}_{\rm int} \rangle + \langle \gamma^{(o)}_{\rm int} \gamma^{(r)}_{\rm int} \rangle + \langle \gamma^{(o)}_{\rm sys} \gamma^{(r)}_{\rm sys} \rangle
\end{equation}
The second and third terms are the GI systematic alignment, which still survives. However, the fourth term involves the correlation between optical and radio 
shapes, which will be less than that between one frequency alone as the emission mechanisms are different (c.f. \cite{patel10} 
where no correlation 
at zero lag was found). This term is therefore reduced. Most importantly, the fifth term involving systematics is expected to be zero, as the 
systematics in these two telescopes of completely different design and function are not expected to be correlated at all. We are therefore able 
to remove the dangerous systematic correlations from our shear analysis - and to gain an estimate of its magnitude in the autocorrelation case.

\section{Simulation Synergies}
Both Euclid and the SKA will require a large number of N-body simulations in order to achieve their science objectives. 
These simulations will be used for a variety of purposes for example a small number of large box-size simulations can 
be used to generate mock catalogues that can be used in the development of data analysis algorithms by creating 
fake data with which to test the pipelines. In order to compute the likelihood functions used for cosmological 
analyses covariance matrices for data vectors will be needed. In order to characterise these 
realisations of data will be required \citep{taylor12}, and depending on the number of data points and 
parameters this could result in up to $\sim 10^6$ realisations. Because the covariance matrices for cosmological 
observable power spectra depend on cosmology themselves this results in the potential need for tens of thousands or 
millions of high resolution, small box-size, simulations. 

This is a significant computational task and, because at 
small scales the impact of baryonic feedback will be important (see for example \cite{tdk14}), it is likely that 
these simulations will need to be hydrodynamical as well as N-body. The completion of this number of simulations by 2020, 
in time for both Euclid and the SKA, is a challenge but because the area and depth of both surveys is synergistic it is 
likely that well designed simulations will meet the needs of both experiments, providing a common resource for both. 
Whilst common simulations will reduce the overall resources required for this task, it will also require that these 
simulations are potentially more sophisticated than survey-specific ones. For example a single simulation that provided 
a resource for both 21cm mapping and lower redshift cosmic shear may need to perform ray tracing over a substantially larger redshift 
range than a simulation that was tailored to either cosmological probe alone. 

\section{Methodological Synergies}
Not only will Euclid and the SKA have many scientific synergies, but
they will both provide `big' (a large number of bytes) astrophysical data-sets, the analysis of
which will also present many methodological synergies.  In addition to
the techniques already in use today to analyse optical and radio
observations, recent experience from the analysis of cosmic microwave
background (CMB) observations can be used to infer future methods of
interest.

Bayesian statistics is the dominant paradigm for the analysis of CMB
observations.  Typically cosmological parameters are estimated from
observational data in a Bayesian framework using Markov Chain Monte
Carlo (MCMC) sampling techniques e.g. \cite[e.g.][]{lewis:2002}, 
such as
the Metropolis Hastings algorithm.  If the parameter space is of
moderate size, then nested sampling methods 
\citep{skilling:2004,feroz:multinest1, feroz:multinest2, feroz:multinest3} 
are often used to enable the efficient calculation
of the Bayesian evidence in order to distinguish between cosmological
models.  Such techniques are already common when analysing optical and
radio observations.  For high-dimensional problems, Gibbs and
Hamiltonian sampling have been applied successfully to sample
high-dimensional posterior distributions for CMB analysis \citep[e.g.][]{wandelt:2004, taylor:2008}.  
These sampling techniques are already starting to be applied to mock galaxy surveys \citep{jasche:2010,jasche:2013a,jasche:2013b}.

In addition to Bayesian analysis, sparsity-based approaches are
gaining popularity as an alternative and complementary CMB analysis
technique.  First works in this area focused on the use of wavelets,
which are a powerful signal analysis tool due to their ability to
represent signal content in space and scale simultaneously.  This
property is particularly useful for the study of astrophysical
signals, where the physical processes responsible for signals of
interest are typically manifest on specific physical scales, while
often also spatially localised.  Wavelet transforms and fast
algorithms defined on the sphere 
\citep[e.g.][]{antoine:1999, wiaux:2005, mcewen:2006:cswt2, starck:2006, mcewen:2006:fcswt, wiaux:2007:sdw, marinucci:2008, mcewen:2013:waveletsxv, leistedt:s2let_axisym} 
were developed for the analysis of
CMB data and have proved to be particularly effective \citealt{mcewen:2006:review}.  
More recently, the
revolutionary new paradigm of compressive sensing has been exploited
for CMB analysis.  Compressive sensing \citep{candes:2006a,donoho:2006} 
is a recent ground-breaking
development in information theory, which goes beyond the
Shannon-Nyquist sampling theorem by exploiting sparsity and which has
the potential to revolutionise data acquisition in many fields (for a
brief introduction see \citealt{baraniuk:2007}).  At its heart,
compressive sensing provides a powerful framework for solving inverse
problems, typically exploiting the sparse representation of signals in
a wavelet dictionary.  Sparse component separation techniques that
exploit these ideas have been developed to recover CMB maps from
observations corrupted by foreground emission \citep{bobin:2013}. There is considerable scope to extend sparsity-based approaches to the
analysis of both Euclid and SKA data, leading to numerous
methodological synergies.

The galaxies that will be observed by Euclid live in 3D, hence to
exploit sparsity-based techniques for the analysis of Euclid data
suitable 3D wavelet transforms must be constructed.  Recently, wavelet
transforms have been constructed specifically for the analysis of such
data by \citet{lanusse:2012} and \citet{leistedt:flaglets}.  These
constructions are complementary, where the construction of \citet{lanusse:2012} 
yields isotropic 3D wavelets, whereas the
construction of \citet{leistedt:flaglets} 
yields wavelets with angular
and radial aperture that can be controlled separately.  Application of
these wavelet transforms to galaxy surveys is underway.  Specifically,
such wavelet representations should be particularly useful for 3D weak
lensing.  For example, compressive sensing approaches have been
developed already to solve the inverse problem required to recover 3D
density maps from weak lensing data using hybrid 2D-1D wavelets \citep{leonard:2012,leonard:2013}.

Sparsity-based approaches have also shown considerable promise for
imaging observations from radio interferometric telescopes, such as
the SKA.  Compressive sensing approaches have been developed to solve
the inverse problem of recovering images from the raw Fourier
measurements made by interferometric telescopes, starting from
idealised applications \citep{wiaux:2009:cs,mcewen:riwfov}, to
optimising telescope configurations \citep{wiaux:2009:ss}, to novel
imaging methodologies \citep{carrillo:sara}, to realistic
imaging scenarios \citep{carrillo:purify}, and to
supporting wide fields-of-view \citep{mcewen:riwfov,wolz:spread_spectrum}. In addition to imaging raw
data, methods to exploit sparsity have been developed to separate the
21cm emission from the epoch of reionization from foreground emission \citep{chapman:2013}.

Sparsity and compressive sensing hence show considerable potential for
the analysis of both Euclid and SKA observations.  Although
appropriate wavelets must be constructed on the space where the data
live, the application of wavelet and compressive sensing techniques to
both types of observations shares many similarities.  A generic
approach can be taken to solve inverse problems in a compressive
sensing framework across different data domains.  Furthermore, the
underlying convex optimisation algorithms e.g. \cite{combettes:2011} 
developed to solve these inverse
problems are largely the same and hence the same underlying codes can
be used or adapted.

In summary, the analysis of the large data-sets expected from both
Euclid and the SKA will exhibit many methodological synergies and will
benefit considerably by the sharing of expertise, techniques, and
numerical codes, both for Bayesian and sparsity-based approaches.

\section{Conclusion} 
The concurrent development of Euclid and the SKA over the next decade, with 
the near simultaneous observations of significant fractions of the sky 
in radio wavelengths and spaced-based optical and NIR imaging and spectra 
represents an unprecedented era in astronomy; and step-change in data 
quantity and quality that is unlikely to be surpassed in the foreseeable future. 
The synergies between these projects is manifest, and it almost goes without saying that 
the combination of the science from these projects will result in 
more than the sum of the parts. 

The synergies are many-fold enabling the 
experiments to mutually reduce systematic effects; improve statistical constraints 
on parameters, in particular for cosmology; and enable new science. 
In addition there are synergies on the preparatory aspects, in 
particular on the production of simulations, and on the exploitation of the 
data through methodological overlap. Specifically 
\begin{itemize}
\item
{\bf Survey synergies.} Euclid and SKA will observe the same sky, at complementary overlapping depths, 
over the same time period of observation, but crucially at different wavelengths which enables independent 
cross-validation of results and systematics. 
\item
{\bf Cosmology synergies.} The combination of the two experiments can improve cosmological constraints through the 
simple addition of the data, but extra information from the cross-correlation is likely to allow for further gains.
In particular dark energy, modified gravity and non-Gaussianity measurement may be improved significantly through the 
combination of Euclid and the SKA.
\item 
{\bf Systematic synergies.} Euclid and SKA will measure some of the same properties from the same galaxies but at 
at different wavelengths and with different instrumental systematics. By cross-correlating it may be possible 
to construct estimators that have much lower (or even zero) sensitivity to particular systematics, for example 
those associated with galaxy weak lensing measurements.
\item
{\bf Simulations \& Methodological synergies.} Both Euclid and SKA will require many N-body simulations, and new 
methodological techniques, to support the theoretical 
interpretation of the observations. Given the finite resources available to the astronomy community there is an opportunity 
for cooperation in this area during and after the construction phases.
\end{itemize}

These synergies exist \emph{now}, even in pre-construction, during preparation, and 
will continue to grow as data is accumulated. These synergies 
will last for decades after data is collected, involving thousands of scientists world 
wide in mutual collaboration. 

\section*{Acknowledgments}
TDK is supported by a Royal Society University Research Fellowship. 
MLB is supported by the an ERC Starting Grant (Grant no. 280127) and by a STFC Advanced/Halliday fellowship.
RS acknowledges support from both ``ASI I/023/12/0'' and ``MIUR PRIN 2010''. DB is supported by UK STFC grant ST/K00090X/1.

\bibliographystyle{apj}
\bibliography{euclid_ska}

\end{document}